%% file: main.tex
\documentclass[10pt]{article} 
\usepackage{iclr2023_conference,times}
\iclrfinalcopy 

\input{math_commands.tex}

\usepackage{hyperref}
\usepackage{url}
\usepackage{booktabs}
\usepackage{graphicx}
\usepackage{multirow}
\usepackage{pifont}%
\usepackage{listings}
\usepackage{tabularx}
\usepackage{xparse}
\NewDocumentCommand\emojisanta{}{
        \includegraphics[scale=0.24]{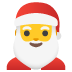}
}
\NewDocumentCommand\emojistar{}{
        \includegraphics[scale=0.24]{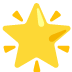}
}
\newcommand{\cmark}{\ding{51}}
\newcommand{\xmark}{\ding{55}}%

\title{\emojisanta{}SantaCoder: don't reach for the stars!\emojistar}

\author{Loubna Ben Allal*\\
Hugging Face\\
\And 
Raymond Li*\\
ServiceNow Research\\
\And 
Denis Kocetkov*\\
ServiceNow Research
\AND
Chenghao Mou\\
Independent\\
\And 
Christopher Akiki\\
ScaDS.AI Leipzig \\ and C4AI Community\\
\And 
Carlos Munoz Ferrandis\\
Hugging Face
\AND
Niklas Muennighoff\\
Hugging Face\\
\And
Mayank Mishra\\
IBM Research\\
\And
Alex Gu\\
MIT\\
\AND
Manan Dey\\
SAP\\
\And
Logesh Kumar Umapathi\\
Saama Technologies\\
\And
Carolyn Jane Anderson\\
Wellesley College\\
\AND
Yangtian Zi\\
Northeastern University\\
\And
Joel Lamy Poirier\\
ServiceNow Research\\
\And
Hailey Schoelkopf\\
EleutherAI\\
\AND
Sergey Troshin\\
University of Amsterdam\\
\And
Dmitry Abulkhanov\\
Huawei Noah's Ark Lab\\
\And
Manuel Romero\\
Independent\\
\AND
Michael Lappert\\
Berner Fachhochschule\\
\And
Francesco De Toni\\
UWA\\
\And
Bernardo García del Río\\
Flowrite\\
\AND
Qian Liu\\
Sea AI Lab
\And
Shamik Bose\\
Independent\\
\And
Urvashi Bhattacharyya\\
Discover Dollar Pvt Ltd\\
\AND
Terry Yue Zhuo\\
CSIRO's Data61 and Monash University\\
\And
Ian Yu\\
PIISA
\And
Paulo Villegas\\
Telefonica I+D\\
\AND
Marco Zocca\\
Unfold ML\\
\And 
Sourab Mangrulkar\\
Hugging Face\\
\And
David Lansky\\
Independent\\
\AND
Huu Nguyen\\
Ontocord, LLC
\And
Danish Contractor\\
Independent
\And
Luis Villa\\
Independent
\AND
Jia Li\\
Independent\\
\And
Dzmitry Bahdanau\\
ServiceNow Research\\
\And
Yacine Jernite\\
Hugging Face
\AND
Sean Hughes\\
ServiceNow\\
\And
Daniel Fried\\
Carnegie Mellon University\\
\And
Arjun Guha\\
Northeastern University and Roblox\\
\AND
Harm de Vries‡\\
ServiceNow Research\\
\And
Leandro von Werra‡\thanks{Corresponding authors (denoted by ‡) can be contacted at \url{contact@bigcode-project.org}}\\
Hugging Face\\
}

\begin{document}

\maketitle

\begin{abstract}
The BigCode project is an open-scientific collaboration working on the responsible development of large language models for code.\footnote{See \url{https://www.bigcode-project.org}} This tech report describes the progress of the collaboration until December 2022, outlining the current state of the Personally Identifiable Information (PII) redaction pipeline, the experiments conducted to de-risk the model architecture, and the experiments investigating better preprocessing methods for the training data. We train 1.1B parameter models on the Java, JavaScript, and Python subsets of The Stack~\citep{Kocetkov2022TheStack} and evaluate them on the MultiPL-E text-to-code benchmark~\citep{cassano2022multiple}. We find that more aggressive filtering of near-duplicates can further boost performance and, surprisingly, that selecting files from repositories with 5+ GitHub stars deteriorates performance significantly. Our best model outperforms previous open-source multilingual code generation models (InCoder-6.7B and CodeGen-Multi-2.7B) in both left-to-right generation and infilling on the Java, JavaScript, and Python portions of MultiPL-E, despite being a substantially smaller model. All models are released under an OpenRAIL license at \url{https://hf.co/bigcode}. 
\end{abstract}

\section{Introduction}
Over the last two years, we have witnessed tremendous progress in the development of code generating AI assistants~\citep{chen2021codex,chowdhery2022palm,Nijkamp2022ACP,fried2022incoder,li2022competition,mbxp}. Machine learning models are now capable of assisting professional developers through the synthesis of novel code snippets, not only from surrounding code fragments, but also from natural language instructions. The models powering these code completion systems are usually referred to as Large Language Models for Code---or code LLMs---and are created by training large transformer neural networks~\citep{vaswani2017attention} on big corpora of source code. However, with the exception of a few small-scale efforts~\citep{xu2022systematicevaluation}, there is generally a lack of transparency on the development of code LLMs, in part due to their commercial value and the legal uncertainty around distributing training data and models. Some groups have released model weights~\citep{fried2022incoder,Nijkamp2022ACP} or provided access to the model through a paid API service~\citep{chen2021codex,mbxp}, but these works did not release the full training data or the preprocessing methods that were used.  

BigCode\footnote{See \url{https://www.bigcode-project.org}} is an open scientific collaboration working on the responsible development of large language models for code, empowering the machine learning and open-source communities through open governance. BigCode was inspired by the BigScience project, an open-scientific collaboration which culminated in July 2022 with the release of a large multi-lingual language model~\citep{scao2022bloom}. As in BigScience, various BigCode working groups focus on relevant subtopics such as collecting datasets, implementing methods for training code LLMs, developing an evaluation suite, and discussing ethical best practices for these powerful models. For example, the Legal, Ethics, and Governance working group has explored questions on data licensing, attribution of generated code to original code, the redaction of Personally Identifiable Information~(PII), and the risks of outputting malicious code. In earlier work, the BigCode community released The Stack v1.1~\citep{Kocetkov2022TheStack}, a 6.4~TB dataset of permissively licensed source code in 384 programming languages. That work also introduced ``Am I in The Stack'',\footnote{\url{https://huggingface.co/spaces/bigcode/in-the-stack}} a governance tool for developers to check whether their source is part of the dataset, and an opt-out form for those who wish to have their code removed from the dataset.\footnote{\url{https://www.bigcode-project.org/docs/about/the-stack/}} 

In this tech report, we summarize the learnings of the BigCode community in developing the Santa models, a set of 1.1B-parameter models trained on the Java, JavaScript, and Python subsets of The Stack and evaluated on MultiPL-E~\citep{cassano2022multiple}. We describe the first steps of the community towards developing larger code models and report experiments to de-risk the model architecture and the data processing pipeline. Specifically, the contributions of this report can be summarized as follows:
\begin{itemize}
    \item We describe the current state of the PII redaction pipeline. We detail how we create a PII benchmark of 400 code files, describe the filters for detecting emails, ip addresses, and secret keys, and analyze its performance on the annotation benchmark. All experiments in this work are conducted on the PII-redacted version of The Stack.  
    \item We run ablations for Multi Query Attention (MQA)~\citep{shazeer2019mqa,chowdhery2022palm,li2022competition} and Fill-in-the-Middle (FIM)~\citep{fried2022incoder,bavarian2022fim}. MQA can significantly speed-up inference for larger batch sizes, while FIM enables code models to do infilling tasks. We find that both changes only slightly deteriorate downstream performance compared to baseline models. 
    \item We investigate the impact of 4 preprocessing methods on the training data: filtering files from repositories with 5+ GitHub stars, filtering files with a high comments-to-code ratio, more aggressive filtering of near-duplicates, and filtering files with a low character-to-token ratio. We observe modest impact of the new filters except for the stars filter, which deteriorates performance on text2code benchmarks significantly. This is an interesting result given that previous work has explicitly filtered for GitHub Stars as a proxy for data quality~\citep{gao2020pile,xu2022systematicevaluation}. 
    \item Using the findings from these experiments, we train a final 1.1B parameter model, dubbed SantaCoder, on Python, JavaScript, and Java. This model obtains comparable or stronger performance than previous open-source multilingual models, InCoder-6.7B and CodeGen-Multi-2.7B, on code generation and infilling tasks on the MultiPL-E benchmark for these three languages, despite being substantially smaller.
\end{itemize}

\section{Related Work}

\paragraph{Code LLMs}
Recently, there has been an increasing amount of research on using large-scale transformer models to analyze or generate source code. Many studies have focused on using decoder-only models with a causal language modeling objective~\citep{chen2021codex,austin2021program,Nijkamp2022ACP,christopolou2022pangucoder,izadi2022codefill,xu2022systematicevaluation,mbxp}, while other studies have investigated encoder~\citep{feng-etal-2020-codebert,kanade2020embeddings} and encoder-decoder architectures~~\citep{li2022competition,ahmad-etal-2021-unified,wang-etal-2021-codet5,roziere2021dobf}. \citet{bavarian2022fim,fried2022incoder} propose to use decoder-only models for code-infilling tasks using a causal masking mechanism, and \citet{bavarian2022fim} argues that training with such a fill-in-the middle (FIM) objective does not harm the model's ability to do left-to-right generation. \citet{shazeer2019mqa} proposes Multi Query Attention (MQA), an architectural change to the transformer neural network in which key and value embeddings are shared across attention heads, resulting in lower memory requirements and faster inference for large batch settings. Multi Query Attention was implemented in AlphaCode~\citep{li2022competition} and PaLM~\citep{chowdhery2022palm}.  


\paragraph{Evaluating text-to-code}
The correctness of generated code can be tested using \textit{unit tests}, a method for approximating semantic equivalence. Textual similarity metrics have also been used to evaluate code \citep{feng2020codebert,ren2020codebleu}; however, they have been shown to correlate only weakly with code correctness \citep{austin2021program,chen2021codex}.

Many single-language benchmarks for evaluating code completion exist \citep{kulal2019spoc,iyer2018mapping,wikisql,spidersql,austin2021program,hendrycks2021measuring,chen2021codex,austin2021program,mbxp,Lai2022DS1000}. Two of the most popular benchmarks for Python are HumanEval \citep{chen2021codex} and MBPP \citep{austin2021program}, which consist of a natural language description of a function and a set of unit tests.

MultiPL-E \citep{cassano2022multiple} extends two popular benchmarks for code completion, MBPP and HumanEval, to 18 additional languages. The doctests, function signatures, and unit tests for each benchmark suite are automatically compiled to new languages. Python-specific terminology in the prompt is automatically replaced with the equivalent terminology used for each programming language.
MBXP~\citep{mbxp} is a concurrent benchmark that uses a similar approach, but differs in the details of type inference, prompt construction, and evaluation. In particular, MBXP uses the same set of assertions in the prompt that it uses to test the correctness of generated solutions. In contrast, MultiPL-E keeps the tests hidden from the model and only uses them to test correctness.

\paragraph{Evaluating other tasks}

Code generation models have also been used to solve a variety of tasks \citep{tufano2020unit,feng2020codebert,ahmed2022multilingual,hellendoorn:dl-ti,pradel:typewriter}. CodeXGLUE~\citep{lu2021codexglue} is a set of 14 datasets for evaluating code generation models. The tasks include code-to-code tasks like clone detection, code repair, and code translation; text-to-code tasks like code search and code generation; and code-to-text tasks like generating documentation. The programming languages included vary by task; the most common are Python and Java. 

\section{Opt-out process}
Developers who do not wish their source code to be used for training code LLMs are given the opportunity to opt-out of The Stack~\citep{Kocetkov2022TheStack}.  We received 9 opt-out requests before the cut-off date for removing data (31 October 2022). These individuals accounted for 299 repositories. Of these, 161 were already excluded from The Stack v1.0 (because they did not have a permissive license), and 138 were in The Stack v1.0. We honored the requests to opt-out and removed these repositories from The Stack v1.1. After the cut-off date (31 October 2022), we have received more requests for requests and we will remove these repositories prior to releasing The Stack v1.2.


\section{Redacting Personally Identifiable Information }\label{sec:PII}
We describe our first efforts to redact PII from The Stack. 

\subsection{PII benchmark}
We construct a PII benchmark by
annotating the following entities on a small subset of The Stack: names, emails, usernames, passwords, IP addresses, API keys, and SSH keys. We pre-filtered 400 samples from a total of 4000 code files that were likely to contain Personally Identifiable Information (PII). We first select 4000 code files from 11 programming languages, with a total of 800 samples for Python and C++, 400 samples for Java, JavaScript, TypeScript, and PHP, and 160 samples for C, C\#, Markdown, Go, and Ruby. To detect keys in these samples, we used the detect-secrets tool\footnote{\url{https://github.com/Yelp/detect-secrets}} with all default plugins activated. In addition, we used regular expressions to detect emails, IPv4 and IPv6 addresses, see Appendix \ref{sec:pii_regex}. Twelve members of the BigCode community annotated the files on the LightTag platform\footnote{\url{https://www.lighttag.io/}}, with one annotator assigned per file. After the annotation phase, one member reviewed all the annotation tags. To further increase annotation quality, we ran our initial PII detection tools on the annotated files and manually corrected any incorrect annotations identified as false positives or false negatives.


\subsection{PII detection and redaction}
For the first iteration of the PII redaction pipeline, we focus on emails, IP addresses, and keys, and leave the detection of names, usernames, and passwords for future work.  

\paragraph{Emails} We use a regular expression to detect emails, see Appendix \ref{sec:pii_regex}. We replace detected emails with [random 5 character string]@example.com. 

\paragraph{IP addresses} We use regular expressions for IPv4 and IPv6 IP addresses, see Appendix \ref{sec:pii_regex}. In addition, we check if the detected IP addresses have a valid format using the \texttt{ipaddress} python package. We also do not select IP addresses of the format a.b.c.d where a, b, c and d are single digit numbers, except if the words “dns” or “server” appear in the neighboring context (100 characters before or after). These detected addresses were mostly false positives, consisting of package and release versions. Lastly, we do not anonymize private IP addresses\footnote{They are non-internet facing IP addresses used in internal networks} and popular DNS servers, as we don’t consider them sensitive information. See Appendix \ref{sec:privateIPaddress} for the full list. 

We replace detected IP addresses with one of 5 randomly generated IP addresses. 

\paragraph{Keys} We employed the \texttt{detect-secrets} tool to identify secret keys in the code files. To this end, we kept all the regex and entropy based plugins, including the AWS key detector, the GitHub Token detector, the Azure storage key detector, and the Base64 High Entropy String detector.
You can find the full list of plugins in Appendix \ref{sec:detect-secrets-plugins}. We deactivated keyword detectors because they were detecting commonly used words like "password" rather than actual secret keys. To remove false positives, we activated filters like UUIDs and string-like secret filtering, see the full list in Appendix \ref{sec:detect-secrets-filters}. We also observed that entropy detectors sometimes detected human-readable text like paths and URLs as secrets, even when adjusting the entropy threshold. To address this issue, we added a \texttt{gibberish}\footnote{\url{https://github.com/domanchi/gibberish-detector}} detector filter on top of detect-secrets to verify that the detected string was actually gibberish. Additionally, we noticed that hashes were sometimes falsely detected as secret keys. To mitigate this problem, we added a hash filter that verifies the size of the detected string and checks for the presence of keywords like ``sha'', ``md5'', ``hash'', and ``byte'' in the neighboring context. Finally, to avoid corrupting any files, we prevent the removal of keys from files where words like ``sha'' or ``hash'' are mentioned in more than 2\% of the number of lines.

\subsection{Performance analysis}
\begin{figure}[t] 
\centering
   \begin{minipage}[t]{0.40\linewidth}
      \centering \includegraphics[width=5.5cm, height=5cm]{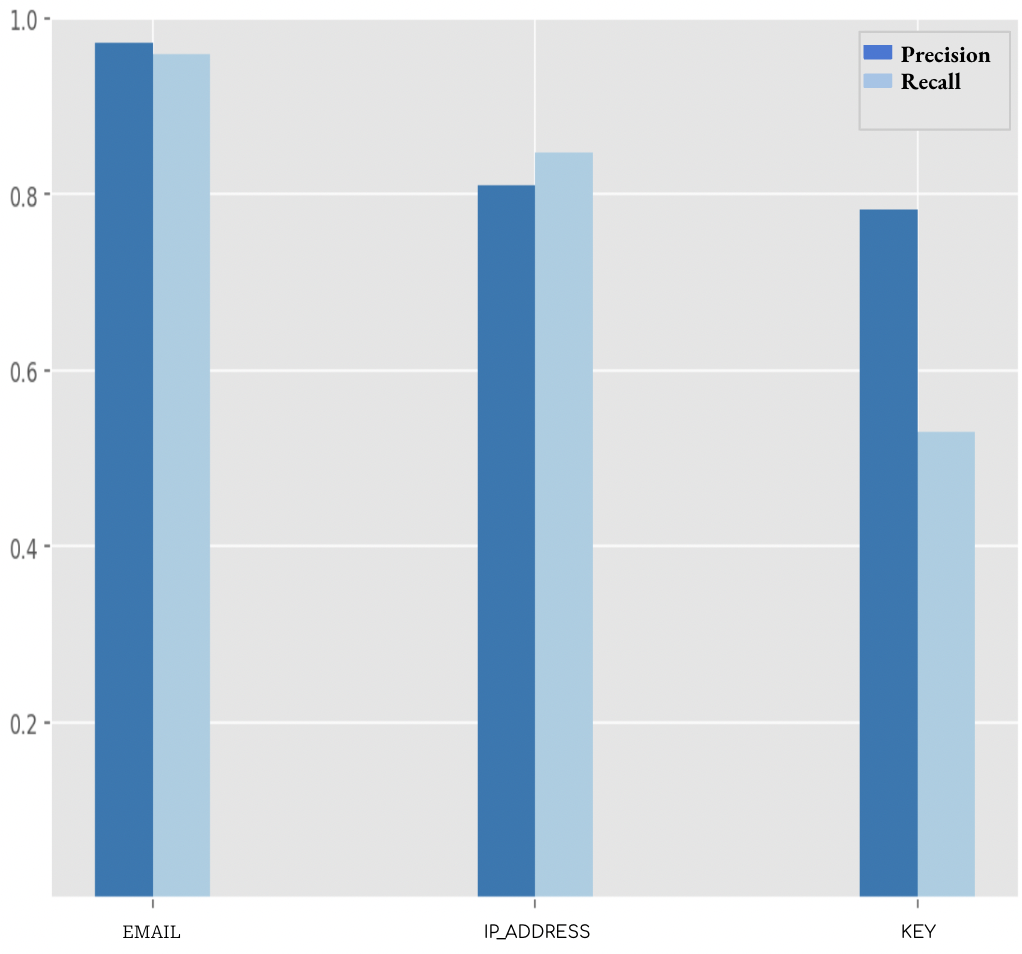}
      \caption{Precision and recall of PII detection tools.}
      \label{fig:precision-recall} 
   \end{minipage}\hspace{15mm}
   \begin{minipage}[t]{0.40\linewidth}
      \centering \includegraphics[width=5.5cm, height=5cm]{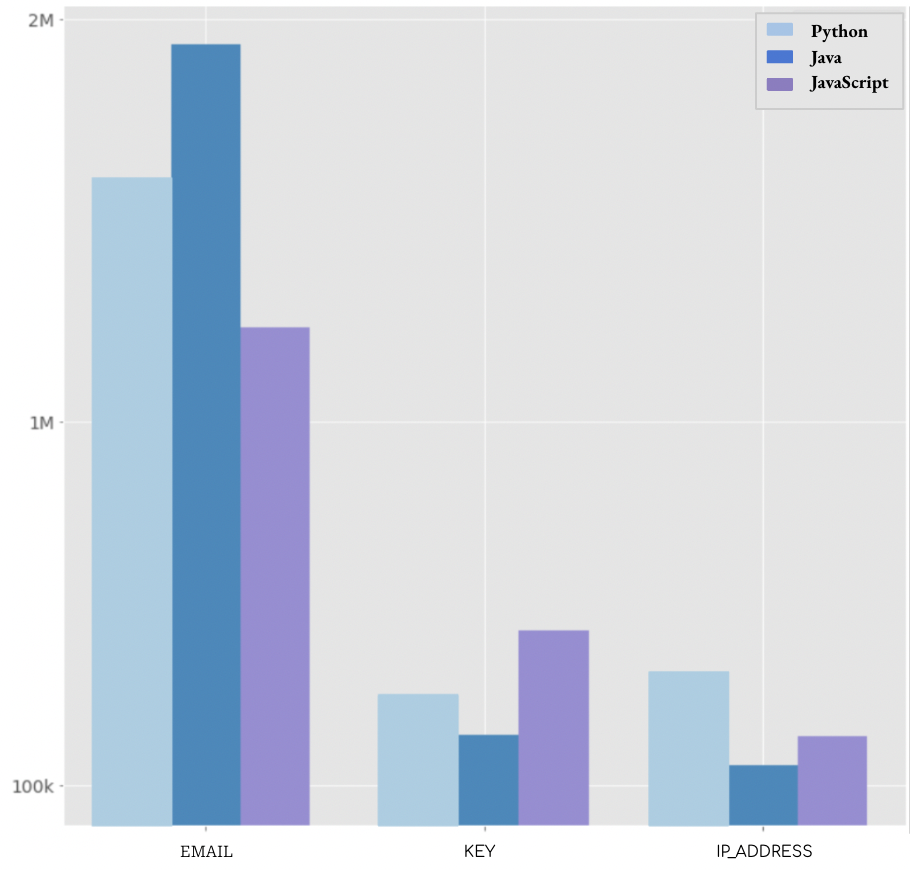}
      \caption{Distribution of PII detected in The Stack for Python, Java and JavaScript.}
      \label{fig:secrets-dist} 
   \end{minipage}
\end{figure}

\paragraph{Evaluation on PII benchmark} We evaluated our PII detection pipeline on the benchmark we annotated. The 400 files contained 214 emails, 99 IP addresses and 34 secret keys. Figure \ref{fig:precision-recall} shows the precision and recall for each PII entity. 
Email and IP address detection perform well, with a precision and recall above 90\% for emails and above 80\% for IP addresses. While key detection also achieves almost 80\% precision, its recall is much lower (slightly above 50\%). We found that the key detection pipeline was especially sensitive to the precision-recall trade-off, as including more plugins or disabling some filters detected more keys but also increased the number of false positives.

\paragraph{PII detection on The Stack}
We run the PII pipeline on the Python, Java and JavaScript subsets of The Stack v1.1~\citep{Kocetkov2022TheStack}. Table \ref{tab:pii-stack-numbers} shows some statistics on the number of files containing PII and the total number of secrets found. Some files containing PII are not modified if they contain only private IP addresses or popular DNS servers, as explained in the previous section. The number of files containing PII is significantly lower for JavaScript compared to Python and Java, but this could be due to the fact that JavaScript files were filtered based on line length and percentage of alphanumeric characters before running PII detection. We also observe that Python and JavaScript have a higher number of secrets per file compared to Java.

To better understand these results, we computed the relevant percentiles in Table \ref{tab:pii-stack-stats}. We can see that Java indeed has fewer secrets per file, and that almost 0.1\% of the files contain a large number of secrets (about 100). We found that some of these files contained multiple instances of PII, such as emails stored in some form of database, or are files containing long encodings and key-like strings that are split into multiple keys. Finally, we also plot the distributions of detected secrets by entity type in Figure \ref{fig:secrets-dist}. For this graph, we filtered out files with more than 100 secrets, but this did not change the distribution of PII across languages. We observe that IP addresses are most often found in Python, keys in JavaScript and emails in Java.

\begin{table}[t]
\centering
\begin{tabular}{lcccc}
\toprule
Language & \# files & \# files with PII  & \# secrets & \# modified files\\
\midrule
Python & 15,148,604 & 1,224,632 & 3,255,053 & 1,040,809\\
Java & 25,124,914 & 1,588,453 & 2,757,169 & 1,506,766 \\
JavaScript* & 23,670,848 &  \phantom{0,}835,198 & 2,468,183 & \phantom{0,}744,842\\
\bottomrule
\end{tabular}
\caption{Statistics from running PII detection on The Stack. JavaScript files initially went through line-length filtering. Modified files are those altered during PII redaction.} 
\label{tab:pii-stack-numbers}
\end{table}

\begin{table}[t]
\centering
\begin{tabular}{cccccc}
\toprule
Language & mean & median & 95th percentile  & 99th percentile & 99.9th percentile\\
\midrule
Python & 2.7 &1 & 6 & 23 & 135\\
Java & 1.7 &1 & 3 & 11 & 63 \\
JavaScript & 3.3 &1 & 7 & 30 & 197\\
\bottomrule
\end{tabular}
\caption{Statistics of the number of detected PII per file in The Stack.} 
\label{tab:pii-stack-stats}
\end{table}


\section{Experiments}

\subsection{Dataset, model, and training details}\label{sec:dataset_model}
\paragraph{Dataset} The base training dataset for the experiments in this paper contains 268 GB of Python, Java and JavaScript files from The Stack v1.1 \citep{Kocetkov2022TheStack} after removing data from opt-out requests, near-deduplication, PII-redaction (see Section \ref{sec:PII}), and filtering based on line-length and percentage of alphanumeric characters. This dataset was also decontaminated by removing files that contained test-samples from the following benchmarks: HumanEval~\citep{chen2021codex}, APPS~\citep{hendrycks2021measuring}, MBPP~\citep{austin2021program} and MultiPL-E~\citep{cassano2022multiple}.

\paragraph{Tokenizer}
Seeing as the Santa models were the first models our community would train, our design choices for the tokenizer were modulated by a conservative approach, partly based on insights developed during the development of InCoder~\citep{fried2022incoder}. We train a Hugging Face Tokenizer~\citep{anthony_moi_2022_hftokenizers} using the Byte-Pair Encoding~(BPE) algorithm on raw bytes with a vocabulary size of 49,152 tokens. This tokenizer was trained on 600,000 rows (Around 2.6 GB) of data---200,000 for each language---which were pre-tokenized using a digit splitter and the default GPT-2 pre-tokenizer regex before being converted to bytes.


\paragraph{Training details} Our base model is a 1.1B-parameter decoder-only transformer with FIM and MQA trained in \texttt{float16}. It has 24 layers, 16 heads and a hidden-size of 2048. The model is trained for 300K iterations with a global batch-size of 192 using Adam \citep{DBLP:journals/corr/KingmaB14} with $\beta_1=0.9$, $\beta_2=0.95$, $\epsilon=10^{-8}$ and a weight-decay of $0.1$. A total of 118B tokens are seen in training. The learning-rate is set to $2\times10^{-4}$ and follows a cosine decay after warming up for 2\% of the training steps.
Each training run takes 3.1 days to complete on 96 Tesla V100 GPUs for a total of $1.05 \times 10^{21}$ FLOPs.
The final model described in Section \ref{sec:final_model} uses twice the amount of compute.

\subsection{Architecture ablations}

We perform ablation experiments to de-risk the model architecture and training objective. Specifically, we investigate Fill-in-the-Middle~\citep{bavarian2022fim} and Multi Query Attention (MQA)~\citep{shazeer2019mqa}.

\paragraph{FIM vs No-FIM}
Recent works \citep{fried2022incoder, bavarian2022fim} have shown that autoregressive language-models can learn to infill code snippets by random transformation of the training data. \citet{bavarian2022fim} argue that such data transformations do not harm the left-to-right generative capabilities of the model. 
Following \citet{bavarian2022fim}, we implement FIM as a random transformation of the input sequence and split each training document into three parts uniformly at random: prefix, middle and suffix.
Each part is prepended with a corresponding sentinel token, then documents are rearranged to put the middle part at the end of the sequence. The autoregressive training objective is unchanged. We use context-level FIM, apply transformations at the character level, use a FIM-rate of $0.5$ and SPM+PSM joint training. We compare our base model to a model that was trained with the standard left-to-right objective only (No-FIM).

\paragraph{Multi Query Attention vs Multi Head Attention}
\citet{shazeer2019mqa} proposes Multi Query Attention (MQA), an architectural change to transformer that shares key and value embeddings across attention heads. Compared to Multi Head Attention (MHA), this lowers the memory bandwidth requirements at generation time and results in faster inference. We compare our base model to a similar model using MHA instead, with the same hyper-parameters otherwise. Note that the MHA model has more parameters (1.3B) than the base model in this setting. 


\subsection{Data filtering ablations}
We experiment with a number of preprocessing methods applied to the base dataset, described in Section \ref{sec:dataset_model}. Note that the filters are applied on top of the other filters such as near-deduplication, line length filtering, etc.  

\begin{table}[t]
\small
\begin{tabular}{lccccc}
\toprule
Language & Base & Stars & Comments-to-code & Near-dedup & Tokenizer fertility\\
\midrule
Python & 75.6 GB & 26.6 GB & 65.6 GB & 62.0 GB & \phantom{0}72.5 GB\\
Java & 110 GB & 35.8 GB & 92.7 GB & 88.4 GB & 105.5 GB \\
JavaScript & 82.7 GB & 20.8 GB & 57.5 GB & 65.1 GB & \phantom{0}76.4 GB\\
\bottomrule
\end{tabular}
\caption{Data volume after additional filtering of the Python, Java, JavaScript subsets of The Stack. } 
\label{tab:sizes}
\end{table}

\paragraph{GitHub stars} Do popular repositories contain good quality code? We use GitHub stars as a proxy metric. We set the minimum threshold to 5 stars, as we believe that a lower number of stars would not be an indicator of popularity. This filter removes more than 60\% of the data (in terms of volume), see Table \ref{tab:sizes}. Note that more than 40\% of the files do not have stars and that setting the threshold to 10 stars would remove an additional 5\% of the data. 

\paragraph{Comment-to-code ratio} Good code should be well documented. With this assumption, we filter files with a high comments-to-code ratio.
We use the \texttt{ast} and \texttt{tokenize} modules to extract docstrings and comments from Python files, and \texttt{Pygments} to extract comments from Java and JavaScript files. We then analyze the comment-to-code character ratio. We find that about 20\% of Python and Java files and 45\% of JavaScript files have no comments. We use a minimum threshold of 1\%, removing an additional 3\% of files in each language. We also find that files with a ratio above 80\% have poor quality, so we filter them out, eliminating 2\% of data in all languages. Overall, this comment-to-code filter removes 20\% of the data in terms of volume.

\paragraph{More near-deduplication} While exact-match deduplication is the most common preprocessing step for code LLMs (see Table \ref{tab:dedup}), \citet{Kocetkov2022TheStack} showed that \emph{near}-deduplication leads to additional performance gains. Their near-deduplication pipeline largely inherited the settings from CodeParrot ~\citep{Tunstall2022-nw}: MinHash~\citep{broder2000identifying} + Locality Sensitive Hashing (LSH) based on \texttt{datasketch}\footnote{\url{https://github.com/ekzhu/datasketch}} with unigrams (non-alphanumeric tokens) and a $0.85$ Jaccard similarity threshold. Additionally, it also recalculates the true unigram Jaccard similarity during the post-processing stage to weed out any false positives. In this paper, we investigate whether different deduplication settings can further improve performance.

To this end, we conduct ablation experiments on a 200K subset of the raw python dataset from the Stack v1.1. We investigate the number of false positives and false negatives by comparing the clustered files with their real Jaccard similarity. We find that: 1) Using unigrams during MinHash calculation leads to many false positives, around 20\% at $0.85$. Increasing the n-gram size reduces false positives, but also increases false negatives. This is an expected trade-off between precision and recall; 2) A lower threshold would cause more documents to be removed at the cost of processing time. In our experiments, we have observed good duplicates occur with a similarity as low as $0.65$, even though the FP and FN rates don't change much.

We find that combining 5-grams and a $0.7$ threshold strikes a good balance between false positives and false negatives while removing an additional 16\%--20\% files. In particular, the increased false negatives occur mostly among documents with lower real Jaccard similarity bounds, whereas documents with higher similarities ($> 0.85$) even have a decreased false negative rate (from 35\% to 24\%). Due to time constraints, we apply such deduplication on the already deduplicated datasets using the Stack v1 hyperparameters. We will refer to the final results as more near-deduplication or near-deduplication alt. 

\begin{table}[t]
\centering
\small
\begin{tabular}{ll}
\toprule
Model & Dataset Deduplication Method \\
\midrule
InCoder~\citet{fried2022incoder} & Exact Match (alphanumeric token sequence)\\
CodeGen~\citep{Nijkamp2022ACP} & Exact Match (sha-256) \\
AlphaCode~\citep{li2022competition} & Exact Match (non-whitespace text) \\
PolyCoder~\citep{10.1145/3520312.3534862} & Exact Match (hash) \\
PaLM Coder~\citep{chowdhery2022palm} & Near-deduplication (Levenshtein distance) \\
CodeParrot~\citep{Tunstall2022-nw} & Near-deduplication (MinHash) \\
Codex~\citep{chen2021codex} & Exact Match ("unique python files") \\
\bottomrule
\end{tabular}
\caption{Various deduplication methods adopted for different model training data.} 
\label{tab:dedup}
\end{table}

Unlike other data preprocessing or filtering techniques that target one document at a time, near-deduplication requires a centralized index that can be prohibitive for large data processing. We have released the deduplication code used in this paper on GitHub\footnote{\url{https://github.com/bigcode-project/bigcode-dataset}} and will release a distributed version soon. For reference, it takes about 10 hours to deduplicate 42 million Java documents using plain multiprocessing while it takes less than 40 minutes in a distributed (but comparable) environment.

\paragraph{Tokenizer fertility} Can we use the tokenizer to remove low-quality files from the dataset? We experiment with filtering files with a low character-to-token ratio\footnote{We slightly abuse the term tokenizer fertility in this work as it usually refers to the average number of subwords per token, where a token is determined by the true tokenizer of the programming language. See e.g. \citep{rust-etal-2021-good}}. For each language, we find that files with a ratio below the 5th percentile are usually of poor quality, but increasing the threshold may eliminate some good-quality files. We therefore set the cutoff value for this ratio to the following values: 2.5 for Python, 2.9 for Java, and 2.6 for JavaScript. This filters out roughly 4\% to 5\% of data. Note that these values depend highly on the tokenizer and the data. This filter may also be biased against files with non-English comments.

\subsection{Evaluation}
\label{sec:evaluation}

\begin{table}[t]
\centering
\begin{tabular}{llccc}
\toprule
Language & Attention & FIM  & HumanEval & MBPP\\
\midrule
\multirow{3}{*}{Java} & Multi Query Attention & \cmark & 0.35 & 0.54\\
& Multi Head Attention & \cmark & 0.36 & 0.55\\
& Multi Query Attention & \xmark &  0.37 & 0.55\\
\midrule
\multirow{3}{*}{JavaScript} & Multi Query Attention & \cmark & 0.33 & 0.64\\
& Multi Head Attention & \cmark & 0.37 &0.67 \\
& Multi Query Attention & \xmark & 0.37 & 0.65 \\
\midrule
\multirow{3}{*}{Python} & Multi Query Attention & \cmark & 0.36 & 0.67\\
& Multi Head Attention & \cmark & 0.38 & 0.70\\
& Multi Query Attention & \xmark & 0.39 & 0.68\\
\bottomrule
\end{tabular}
\caption{Pass@100 results for the architecture ablations on HumanEval and MBPP. }
\label{tab:architecture_ablations}
\end{table}

\begin{table}
\centering
\begin{tabular}{llll}
\toprule
Model                         & Java & JavaScript & Python \\
\midrule
Baseline                      & 0.64 & 0.61 & 0.42 \\
GitHub stars                  & 0.54 & 0.57 & 0.37 \\
Comments-to-code              & 0.62 & 0.59 & 0.44 \\
More near deduplication       & 0.66 & 0.57 & 0.45 \\
Tokenizer fertility          & 0.67 & 0.65 & 0.45 \\
\midrule
Final                         & 0.62 & 0.60 & 0.44 \\
\bottomrule
\end{tabular}
\caption{Fill-in-the-middle results for the data filtering ablations on MultiPL-HumanEval.  Each number reports the fraction of lines where the model exactly reproduces a single line of code that is held out from the body of a function in a held out problem. }
\label{table:fim_humaneval_ablations}
\end{table}
\paragraph{Text2code evaluation}

The text2code task involves generating the body of a function from a prompt that includes a function description, the function signature (its name and arguments), and optionally a handful of example inputs and outputs.
Every problem is accompanied by a set of hidden test cases, which are used to determine if the generated function is correct.
We use the MultiPL-E text2code benchmark~\cite{cassano2022multiple}, which is derived from HumanEval~\cite{chen2021codex} and MBPP~\cite{austin2021program} (the ``sanitized'' subset of MBPP.). Whereas the latter two benchmarks target Python, MultiPL-E has a suite of compilers that translate HumanEval and MBPP to 18 other programming languages. Since our models are only trained on Java, JavaScript, and Python, we only evaluate them on these three languages.

We use the methodology of \citet{chen2021codex} and we calculate pass@$k$ rates for ($k = 1, 10, 100$) for every problem.
Intuitively, pass@1 estimates the likelihood a model will generate a correct solution in a single attempt, whereas pass@10 and pass@100 estimate the likelihood that the model will generate a correct solution given 10 and 100 attempts respectively.
Following the literature, we sample 200 completions at temperatures 0.2 and 0.8 and use 0.2 to estimate pass@1 and 0.8 for pass@10 and pass@100.

\paragraph{Fill-in-the-middle evaluation}
\begin{figure}
    \centering
    \includegraphics[width=0.95\textwidth]{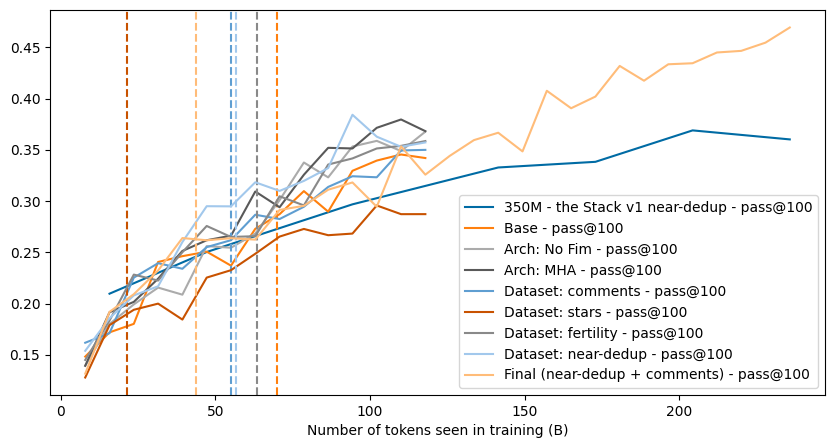}
    \caption{HumanEval pass@100 performance throughout training for all models. Note that evaluation shown here is based on OpenAI Python prompts and might differ (slightly) from the MultiPL-E prompts used in the rest of this paper.}
    \label{fig:human_eval_training}
\end{figure}
To evaluate fill-in-the-middle, we use the single-line exact match metric, which was introduced by \citet{fried2022incoder} and also employed by \citet{bavarian2022fim}.
For every benchmark problem, we mask out a single line of text from the function body (i.e., not from the function description or signature), and prompt the model to fill in that line of code. We exclude blank lines and comments, and count the number of times the model produces exactly the masked out line.
This benchmark requires working solutions for problems, which MultiPL-E does not have. (A text2code benchmark like MultiPL-E only needs hidden tests.) Instead, of writing solutions by hand, we use solutions generated by a code generation model, which is the approach of \citet{mbxp}. Specifically, we use working solutions produced by \texttt{code-davinci-002} at temperature 0.8. Note that this approach does not produce solutions to every problem, since not all problems are solvable. Moreover, for uniformity, we use this approach for Python as well, even though hand-written Python solutions exist for our benchmarks. We only report fill-in-the-middle evaluations for the data filtering ablations. 

\section{Results}

\subsection{Ablations}
\begin{figure}[t]
    \centering
    \includegraphics[width=0.9\textwidth]{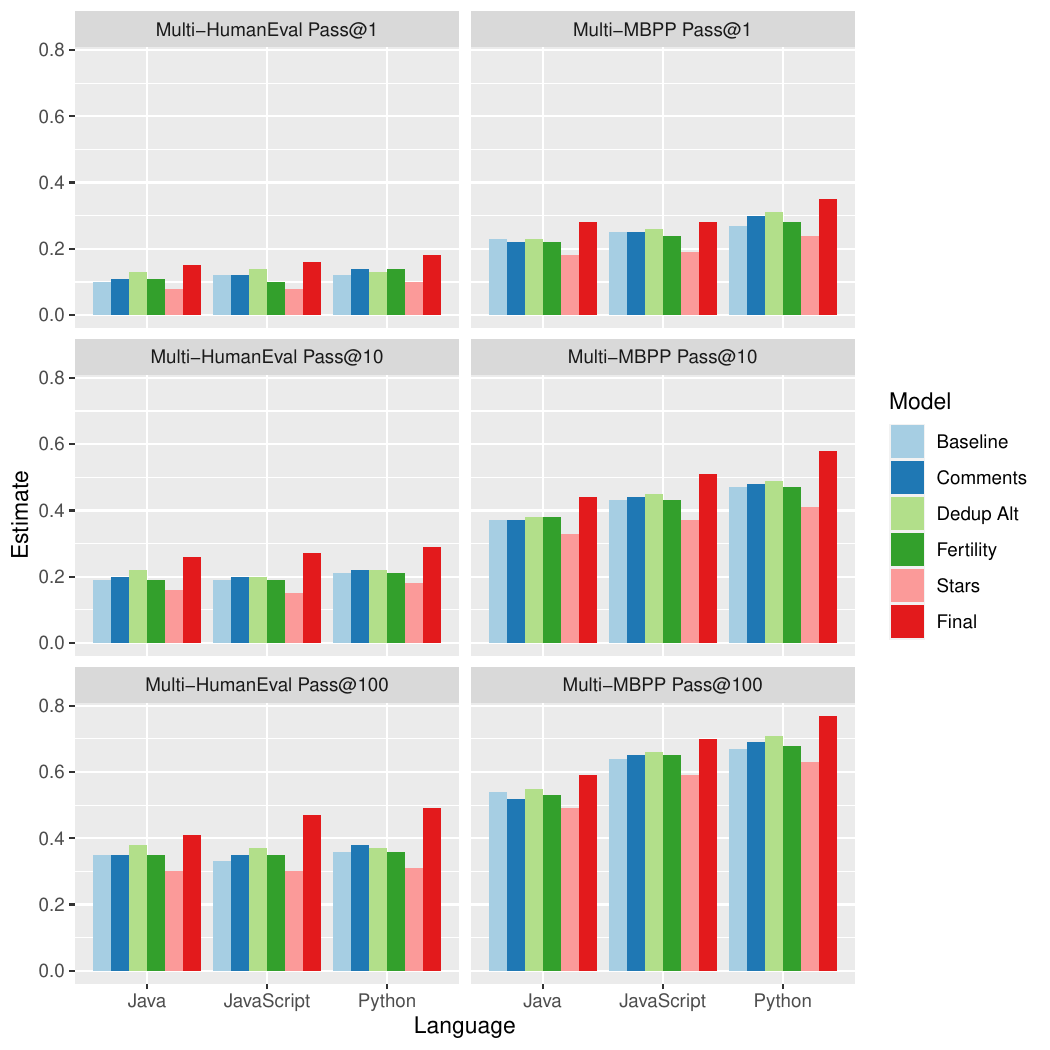}
    \caption{Pass@k rates on Multi-HumanEval and Multi-MBPP by model and language}
    \label{fig:multiple_results}
\end{figure}

For the architecture ablations, we report the results on text2code benchmarks in Table \ref{tab:architecture_ablations}. For the data filtering ablations, we show the text2code results in Figure \ref{fig:multiple_results} and report the fill-in-the middle evaluations in Table \ref{table:fim_humaneval_ablations}.  We show the HumanEval performance throughout all training runs in Figure \ref{fig:human_eval_training}. You can find the full results tables of the text2code experiments are Appendix \ref{sec:text2code_full}. 

\paragraph{Slight drop in performance for MQA} We see a small drop in performance for Multi Query Attention (MQA) compared to Multi Head Attention (MHA). As shown in Table \ref{tab:architecture_ablations}, the MHA model improves pass@100 with 1-4\% on HumanEval and with 1-3\% on MBPP. We specifically observe noticeable improvements for the JavaScript versions of the text2code benchmarks. However, it should be noted that the MHA model has more parameters (1.3B) than the MQA model (1.1B), and a head-to-head comparison might, therefore, not be entirely fair. We think that the inference speed-ups of MQA might outweigh the small drop in performance. 

\paragraph{FIM for cheap} We observe a minor drop in performance of the FIM model compared to the No-FIM model. Specifically, we see that the pass@100 performance of the FIM model is 2-4\% lower on HumanEval and 1\% lower on MBPP. While \citet{bavarian2022fim} presented evidence for the existence of a FIM-for-free property (i.e., arguing that autoregressive models can be trained with FIM without harming left-to-right capabilities), we do find a small but consistent drop of FIM models on left-to-right text2code benchmarks.  

\begin{table}[t]
\centering
\begin{tabular}{ll|ccc|ccc}
\toprule
& & \multicolumn{3}{l}{Left-to-right pass@100} 
 & \multicolumn{3}{c}{Fill-in-the-middle ex.\ match} \\
Model & Size & Java & JavaScript & Python & Java & JavaScript & Python\\
\midrule
InCoder & 6.7B & 0.36 & 0.38 & 0.47 & 0.49 & 0.51 & 0.31 \\
CodeGen-multi & 2.7B & 0.42 & 0.39 & 0.39 & \xmark & \xmark & \xmark \\
CodeGen-mono & 2.7B & \xmark & \xmark & 0.57 & \xmark & \xmark & \xmark \\
Codex\footnotemark
& 2.5B & \xmark & \xmark & 0.60 & \xmark & \xmark & \xmark \\
\midrule
SantaCoder & 1.1B & 0.41 & 0.47 & 0.49  & 0.62 & 0.60 & 0.44 \\
\bottomrule
\end{tabular}
\caption{Comparing the performance of the final version of SantaCoder with InCoder~\citep{fried2022incoder}, CodeGen~\citep{Nijkamp2022ACP}, and Codex~\citep{chen2021codex} on left-to-right (HumanEval pass@100) and fill-in-the-middle benchmarks (HumanEval line filling, exact match).} 
\label{tab:past_work_comparison}
\end{table}
\footnotetext{This is the performance of a Codex model reported by \citet{chen2021codex}. It is not clear if this model is available via the OpenAI API.}

\paragraph{Modest impact of near-deduplication, comments, and fertility filter} On text2code benchmarks, we observe small gains for the near-deduplication and comment-to-code filters and a neutral effect of the tokenizer filter. The near-deduplication filter improves HumanEval performance by 1-3\% and MBPP by 1-4\% across the three programming languages. The comment-to-code filter improves HumanEval performance by 0-2\% but decreases MBPP performance in certain cases (Java). See Appendix \ref{sec:text2code_full} for the full results table. On fill-in-the-middle benchmarks, we see that the tokenizer fertility filter performs well, improving performance by 2-4\% across the three languages. The near-duplication and comments filters have a mixed effect, improving fill-in-the-middle performance for Python but deteriorating performance for JavaScript.

\paragraph{GitHub stars deteriorate performance} Surprisingly, we find that the GitHub stars filter performs poorly. On HumanEval and MBPP, the pass@100 performance consistently drops by 3-6\% across the three languages. On the fill-in-the-middle benchmark, the performance drops by 5-11\% (Table \ref{table:fim_humaneval_ablations}). Note that the stars filter removes the most data (over 60\%) and, therefore, raises the question whether the performance difference is due to the smaller dataset. However, as can be seen in Figure \ref{fig:human_eval_training}, HumanEval pass@100 diverged early on in training, indicating that the drop in performance is not only due to data size but also data quality.

\subsection{Final model}
\label{sec:final_model}
Based on the insights from the architecture and dataset ablations, we train a final model, which we call SantaCoder, with MQA and FIM and the two data filters that yielded the best results: more near-deduplication and comments-to-code filter. We train this model for 600K iterations (236B tokens) and keep all other hyper-parameters the same. 

\paragraph{Improved text2code performance} Doubling the training iterations leads to much stronger text2code performance on MultiPL-E, significantly boosting performance across all benchmarks and programming languages (see Figure \ref{fig:multiple_results}). Looking at the performance throughout training (Figure \ref{fig:human_eval_training}), it is likely that longer training can further increase performance. Surprisingly, we find that the final training run did not improve the fill-in-the-middle evaluations (see Table \ref{table:fim_humaneval_ablations}), at least on these single line infilling tasks.

\paragraph{Comparison to InCoder, CodeGen, and Codex} 

Table~\ref{tab:past_work_comparison} compares our SantaCoder model
to comparably-sized code generation models from previous work on the MultiPL-E benchmark, using the methodology described in  Section~\ref{sec:evaluation}.  
We find that our model generally outperforms previous open-source multi-language code generation models despite being smaller, outperforming the InCoder 6.7B~\citep{fried2022incoder} model on both left-to-right generation and single line fill-in-the-middle infilling across languages, and obtaining comparable or stronger performance to CodeGen-multi 2.7B~\citep{Nijkamp2022ACP}. 




\section{Conclusion}
We described the progress of the BigCode project until December 2022. The community took its first steps towards redacting PII and demonstrated that regular expressions are reasonably effective at detecting emails and IP addresses. Future work should focus on increasing the precision and recall of secret keys, as well as detecting other sensitive information such as names, usernames, and password. Using the PII-redacted version of The Stack, we conducted a series of architectural and data filtering ablations. One of our main findings was that filtering for Github stars consistently decreased performance across all benchmarks and programming languages. Using the findings of these ablation studies, we trained a final 1.1B model---dubbed SantaCoder---for 236B tokens and showed it is able to outperform previous multi-lingual code models (InCoder-6.7B and CodeGen-Multi-2.7B) on both left-to-right generation and infilling tasks. We anticipate that larger architectures and more training data will be able to produce stronger multilingual, infilling-capable models, and plan to continue to scale the findings from our investigations here.

\newpage
\section{Contributions}

\paragraph{Model license}
Carlos Munoz Ferrandis, Christopher Akiki, Danish Contractor, Harm de Vries, Huu Nguyen, Leandro von Werra, Luis Villa, Sean Hughes, Yacine Jernite, David Lansky

\paragraph{PII redaction}
Loubna Ben Allal,  Jia Li, Paulo Villegas, Harm de Vries, Leandro Von Werra, Christopher Akiki, Ian Yu, Michael Lappert, Urvashi Bhattacharyya, Shamik Bose, Bernardo García del Río, Francesco De Toni, Terry Yue Zhuo, Qian Liu, Manuel Romero

\paragraph{Dataset}
Denis Kocetkov, Chenghao Mou, Loubna Ben Allal, Leandro von Werra, Dmitry Abulkhanov, Christopher Akiki, Raymond Li

\paragraph{Tokenizer} Christopher Akiki, Sergey Troshin, Dmitry Abulkhanov, Daniel Fried, Leandro von Werra, Harm de Vries

\paragraph{Training and architecture}
Raymond Li, Daniel Fried, Hailey Schoelkopf, Joel Lamy Poirier, Qian Liu, Niklas Muennighoff, Loubna Ben Allal, Dzmitry Bahdanau, Harm de Vries, Leandro von Werra

\paragraph{Opt out}
Sean Hughes, Carlos Munoz Ferrandis, Christopher Akiki, Denis Kocetkov, Harm de Vries, Huu Nguyen, Leandro von Werra, Luis Villa

\paragraph{Evaluation}
Arjun Guha, Yangtian Zi, Carolyn Jane Anderson, Loubna Ben Allal, Raymond Li, Niklas Muennighoff, Manan Dey, Logesh Kumar Umapathi, Leandro von Werra, Harm de Vries, Marco Zocca

\paragraph{Inference}
Mayank Mishra, Alex Gu, Joel Lamy Poirier, Leandro von Werra, Harm de Vries, Sourab Mangrulka

\paragraph{Acknowledgement}
We thank ServiceNow and HuggingFace for the provided compute resources. 

\bibliography{bigcode}
\bibliographystyle{iclr2023_conference}

\appendix

\section{Full text2code results}\label{sec:text2code_full}
We report the full results of all experiments. Table \ref{table:filtering_humaneval} and \ref{table:filtering_mbpp} show the full results for the data filtering ablations on HumanEval and MBPP, respectively. Table \ref{tab:architecture_humaneval} and \ref{tab:architecture_mbpp} reports the full results for the architecture ablations on HumanEval and MBPP, respectively.  
\begin{table}[htbp]
\centering
\begin{tabular}{lllll}
\toprule
Language & Model & Pass@1 & Pass@10 & Pass@100\\
\midrule
\multirow{5}{*}{Java} & Baseline & 0.1 & 0.19 & 0.35\\
& GitHub stars & 0.08 & 0.16 & 0.3\\
& Comments-to-code ratio & 0.11 & 0.2 & 0.35\\
& More near deduplication & 0.13 & 0.22 & 0.38\\
& Tokenizer fertility & 0.11 & 0.19 & 0.35\\
\midrule
\multirow{5}{*}{JavaScript} & Baseline  & 0.12 & 0.19 & 0.33\\
& GitHub stars & 0.08 & 0.15 & 0.3\\
& Comments-to-code ratio & 0.12 & 0.2 & 0.35\\
& More near deduplication & 0.14 & 0.2 & 0.37\\
& Tokenizer fertility   & 0.1 & 0.19 & 0.35\\
\midrule
\multirow{5}{*}{Python} & Baseline & 0.12 & 0.21 & 0.36\\
& GitHub stars & 0.1 & 0.18 & 0.31\\
& Comments-to-code ratio & 0.14 & 0.22 & 0.38\\
& More near deduplication & 0.13 & 0.22 & 0.37\\
& Tokenizer fertility & 0.14 & 0.21 & 0.36\\
\bottomrule
\end{tabular}
\caption{Full results for data filtering ablations on HumanEval}
\label{table:filtering_humaneval}
\end{table}

\newpage



\begin{table}[htbp]
\centering
\begin{tabular}{lllll}
\toprule
Language & Model & Pass@1 & Pass@10 & Pass@100\\
\midrule
\multirow{5}{*}{Java} & Baseline & 0.23 & 0.37 & 0.54\\
& GitHub stars & 0.18 & 0.33 & 0.49\\
& Comments-to-code ratio & 0.22 & 0.37 & 0.52\\
& More near deduplication & 0.23 & 0.38 & 0.55\\
& Tokenizer fertility & 0.22 & 0.38 & 0.53\\
\midrule
\multirow{5}{*}{JavaScript} & Baseline  & 0.25 & 0.43 & 0.64\\
& GitHub stars & 0.19 & 0.37 & 0.59\\
& Comments-to-code ratio & 0.25 & 0.44 & 0.65\\
& More near deduplication & 0.26 & 0.45 & 0.66\\
& Tokenizer fertility   & 0.24 & 0.43 & 0.65\\
\midrule
\multirow{5}{*}{Python} & Baseline & 0.27 & 0.47 & 0.67\\
& GitHub stars & 0.24 & 0.41 & 0.63\\
& Comments-to-code ratio & 0.3 & 0.48 & 0.69\\
& More near deduplication & 0.31 & 0.49 & 0.71\\
& Tokenizer fertility   & 0.28 & 0.47 & 0.68\\
\bottomrule
\end{tabular}
\caption{Full results for data filtering ablations on MBPP}
\label{table:filtering_mbpp}
\end{table}

\begin{table}[htbp]
\centering
\begin{tabular}{llllll}
\toprule
Language & Attention & FIM  & Pass@1 & Pass@10 & Pass@100\\
\midrule
\multirow{3}{*}{Java} & Multi Query Attention & \cmark & 0.1 & 0.19 & 0.35\\
& Multi Head Attention & \cmark & 0.12 & 0.21 & 0.36\\
& Multi Query Attention & \xmark & 0.11 & 0.21 & 0.37\\
\midrule
\multirow{3}{*}{JavaScript} & Multi Query Attention & \cmark & 0.12 & 0.19 & 0.33\\
& Multi Head Attention & \cmark & 0.13 & 0.21 & 0.37\\
& Multi Query Attention & \xmark & 0.14 & 0.21 & 0.37\\
\midrule
\multirow{3}{*}{Python} & Multi Query Attention & \cmark & 0.12 & 0.21 & 0.36\\
& Multi Head Attention & \cmark & 0.13 & 0.24 & 0.38\\
& Multi Query Attention & \xmark & 0.14 & 0.23 & 0.39\\
\bottomrule
\end{tabular}
\caption{Full results for architecture ablations on HumanEval}
\label{tab:architecture_humaneval}
\end{table}

\begin{table}[t]
\centering
\begin{tabular}{llllll}
\toprule
Language & Attention & FIM  & Pass@1 & Pass@10 & Pass@100\\
\midrule
\multirow{3}{*}{Java} & Multi Query Attention & \cmark & 0.23 & 0.37 & 0.54\\
& Multi Head Attention & \cmark & 0.23 & 0.38 & 0.55\\
& Multi Query Attention & \xmark & 0.23 & 0.37 & 0.55\\
\midrule
\multirow{3}{*}{JavaScript} & Multi Query Attention & \cmark & 0.25 & 0.43 & 0.64\\
& Multi Head Attention & \cmark & 0.26 & 0.46& 0.67\\
& Multi Query Attention & \xmark & 0.23 & 0.44 & 0.65\\
\midrule
\multirow{3}{*}{Python} & Multi Query Attention & \cmark & 0.27 & 0.47 & 0.67\\
& Multi Head Attention & \cmark & 0.31 & 0.49 & 0.7\\
& Multi Query Attention & \xmark & 0.28 & 0.47 & 0.68\\
\bottomrule
\end{tabular}
\caption{Full results for architecture ablations on MBPP}
\label{tab:architecture_mbpp}
\end{table}
\newpage

\section{Docstring generation}
In addition to code completion benchmarks, we also report results on docstring generation. To this end, we evaluate our models on CodeXGLUE code-to-text~\cite{lu2021codexglue}, which is a benchmark constructed from CodeSearchNet~\cite{husain2019codesearchnet}. We use the bigcode-evaluation-harness library~\cite{bigcode-evaluation-harness}, which is derived from lm-evaluation-harness~\cite{eval-harness}. Models are prompted with a Python function signature and asked to output a corresponding docstring. Results are shown in Table \ref{tab:codexglue}. 

\begin{table}[t]
\centering
\begin{tabular}{lll}
\toprule
Model Family & Variant & BLEU\\
\midrule
InCoder & 6.7B & 16.04\\
CodeGen-Mono & 16B & 20.56\\
\midrule
SantaCoder & Baseline & 17.67\\
SantaCoder & No-FIM & 17.71\\
SantaCoder & MHA & 17.72\\
SantaCoder & Bf16 & 17.67\\
\midrule
SantaCoder& GitHub Stars & 18.04\\
SantaCoder & Comments-to-code & 17.81\\
SantaCoder & More near deduplication & 17.65\\
SantaCoder & Tokenizer fertility & 17.64\\
\midrule
SantaCoder & Final & 18.13\\
\bottomrule
\end{tabular}
\caption{CodeXGLUE~\citep{lu2021codexglue} Python Docstring generation smoothed 4-gram BLEU scores using the same methodology as~\citet{fried2022incoder} (L-R single). Models are evaluated zero-shot, greedily and with a maximum generation length of 128.}
\label{tab:codexglue}
\end{table}

\paragraph{Findings} We find all BigCode Santa variants with 1.1B parameters to outperform the 6.7B InCoder model~\citep{fried2022incoder}, which we attribute to differences in the training datasets. Among BigCode models, variants trained on more Python perform better: The \emph{stars} variant with 32\% of Python in its training corpus outperforms the \emph{tokenizer fertility} variant with only 28.5\% of Python (see proportions in Table~\ref{tab:sizes}). The \texttt{bfloat16} is the same as the \emph{no-fim} variant, except for the latter being trained in \texttt{float16}. There's no notable performance difference between the two, likely because at our small scale of 1.1B parameters we did not face any training instabilites.

\paragraph{Qualitative examples} Below is an example prompt from CodeXGLUE. Model generations and the correct solution are in Table \ref{tab:codexgluegen}.
\begin{lstlisting}[language=python]
def dailymotion_download(url, output_dir='.', merge=True, info_only=False, **kwargs):
    """
\end{lstlisting}

\begin{table}[t]
\centering
\begin{tabular}{lll}
\toprule
Model Family & Variant & Generation\\
\midrule
InCoder & 6.7B & Download a video from Dailymotion.\\
CodeGen-Mono & 16B & Downloads Dailymotion videos by URL.\\
\midrule
SantaCoder & Baseline & Download Dailymotion videos.\\
SantaCoder & FIM & Download a video from a dailymotion video.\\
SantaCoder & MHA & Download a video from a Dailymotion video.\\
SantaCoder & bf16 & Download video from dailymotion.com.\\
\midrule
SantaCoder & GitHub stars & Download media from dailymotion.com\\
SantaCoder & Comments-to-code & Download a video from Dailymotion.\\
SantaCoder & More near deduplication & Download a dailymotion video.\\
SantaCoder & Tokenizer fertility & Download a video from Dailymotion.\\
\midrule
Correct solution & & Downloads Dailymotion videos by URL.\\
\bottomrule
\end{tabular}
\caption{CodeXGLUE~\citep{lu2021codexglue} Python Docstring generation examples.}
\label{tab:codexgluegen}
\end{table}


\newpage
\section{PII}\label{sec:pii}
\subsection{Regular expressions}
\label{sec:pii_regex}
\paragraph{Email addresses} We used the following regular expression to detect emails. 
\begin{lstlisting}[language=python] 
email_pattern = r'''
    (?<= ^ | [\b\s@,?!;:)('".\p{Han}<] )
    (
      [^\b\s@?!;,:)('"<]+
      @
      [^\b\s@!?;,/]*
      [^\b\s@?!;,/:)('">.]
      \.
      \p{L} \w{1,}
    )
    (?= $ | [\b\s@,?!;:)('".\p{Han}>] )
'''
\end{lstlisting}

We replace detected emails with [random 5 character string]@example.com. 

\paragraph{IP addresses} We used the following regular expressions to detect IPv4 and IPv6 addresses.

\begin{lstlisting}[language=python] 
ipv4_pattern = r"(?:25[0-5]|2[0-4][0-9]|[01]?[0-9][0-9]?)(?:\.(?:25[0-5]|2[0-4][0-9]|[01]?[0-9][0-9]?)){3}"
ipv6_pattern = r"(?:[0-9a-fA-F]{1,4}:){7,7}[0-9a-fA-F]{1,4}|(?:[0-9a-fA-F]{1,4}:){1,7}:|(?:[0-9a-fA-F]{1,4}:){1,6}:[0-9a-fA-F]{1,4}|(?:[0-9a-fA-F]{1,4}:){1,5}(?::[0-9a-fA-F]{1,4}){1,2}|(?:[0-9a-fA-F]{1,4}:){1,4}(?::[0-9a-fA-F]{1,4}){1,3}|(?:[0-9a-fA-F]{1,4}:){1,3}(?::[0-9a-fA-F]{1,4}){1,4}|(?:[0-9a-fA-F]{1,4}:){1,2}(?::[0-9a-fA-F]{1,4}){1,5}|[0-9a-fA-F]{1,4}:(?:(?::[0-9a-fA-F]{1,4}){1,6})|:(?:(?::[0-9a-fA-F]{1,4}){1,7}|:)|fe80:(?::[0-9a-fA-F]{0,4}){0,4}%[0-9a-zA-Z]{1,}|::(?:ffff(?::0{1,4}){0,1}:){0,1}(?:(?:25[0-5]|(?:2[0-4]|1{0,1}[0-9]){0,1}[0-9])\.){3,3}(?:25[0-5]|(?:2[0-4]|1{0,1}[0-9]){0,1}[0-9])|(?:[0-9a-fA-F]{1,4}:){1,4}:(?:(?:25[0-5]|(?:2[0-4]|1{0,1}[0-9]){0,1}[0-9])\.){3,3}(25[0-5]|(?:2[0-4]|1{0,1}[0-9]){0,1}[0-9])"
ip_pattern = (
    r"(?:^|[\b\s@?,!;:\'\")(.\p{Han}])("
    + r"|".join([ipv4_pattern, ipv6_pattern])
    + ")(?:$|[\s@,?!;:'\"(.\p{Han}])"
)
\end{lstlisting}
\paragraph{Data pre-filtering}
This is the regular expression we used to pre-filter the annotation dataset for data containing emails.
\begin{lstlisting}[language=python] 
email_pattern = r'([^\s@,?!;:\'\"=)(]+@[^,\s!?;,\'\"=]{3,}[\.][^\s\b\'\"@,?!;:)(.]+)'
\end{lstlisting}
For IP addresses, we used the same regular expression as the one used for PII detection.

\subsection{List of private IP addresses and popular DNS servers}\label{sec:privateIPaddress}
\begin{itemize}
\item 8.8.8.8
\item 8.8.4.4
\item 1.1.1.1
\item 1.0.0.1
\item 76.76.19.19
\item 76.223.122.150
\item 9.9.9.9
\item 149.112.112.112
\item 208.67.222.222
\item 208.67.220.220
\item 8.26.56.26
\item 8.20.247.20
\item 94.140.14.14
\item 94.140.15.15
\end{itemize}

\subsection{Detect-secrets filters}\label{sec:detect-secrets-filters}
\begin{itemize}
\item detect\_secrets.filters.heuristic.is\_potential\_uuid
\item detect\_secrets.filters.heuristic.is\_likely\_id\_string
\item detect\_secrets.filters.heuristic.is\_templated\_secret
\item detect\_secrets.filters.heuristic.is\_sequential\_string
\end{itemize}

Implementation available at \url{https://github.com/bigcode-project/bigcode-dataset/blob/6b3f54751b6e38e1ed70f2307331d6943ba39eae/pii/utils/keys_detection.py#L11}. 

\subsection{Detect-secrets plugins}\label{sec:detect-secrets-plugins}

\begin{itemize}
\item ArtifactoryDetector
\item AWSKeyDetector
\item Base64HighEntropyString
\item HexHighEntropyString
\item AzureStorageKeyDetector
\item CloudantDetector
\item DiscordBotTokenDetector
\item GitHubTokenDetector
\item IbmCloudIamDetector
\item IbmCosHmacDetector
\item JwtTokenDetector
\item MailchimpDetector
\item NpmDetector
\item SendGridDetector
\item SlackDetector
\item SoftlayerDetector
\item StripeDetector
\item TwilioKeyDetector
\end{itemize}
Implementation available at \url{https://github.com/bigcode-project/bigcode-dataset/blob/6b3f54751b6e38e1ed70f2307331d6943ba39eae/pii/utils/keys_detection.py#L19}. 

\end{document}

%% file: math_commands.tex
\usepackage{amsmath,amsfonts,bm}

\def\eqref#1{equation~\ref{#1}}

\def\1{\bm{1}}

\DeclareMathAlphabet{\mathsfit}{\encodingdefault}{\sfdefault}{m}{sl}
\SetMathAlphabet{\mathsfit}{bold}{\encodingdefault}{\sfdefault}{bx}{n}

